\documentclass[aps,prb,showpacs,superscriptaddress,amsmath,twocolumn]{revtex4-1}

\usepackage{amsmath}
\usepackage{amssymb}
\usepackage{graphicx}
\usepackage[english]{babel}

\begin{document}

\title{Protected edge states in silicene antidots and dots in magnetic field}

\author{P. Rakyta}
\affiliation{Department of Theoretical Physics, Budapest University of Technology and Economics, H-1111 Budafoki \'ut. 8, Hungary}
\affiliation{MTA-BME Condensed Matter Research Group, Budapest University of Technology and Economics, H-1111 Budafoki \'ut. 8, Hungary}

\author{M. Vigh}
\affiliation{Department of Physics of Complex Systems,E{\"o}tv{\"o}s University,H-1117 Budapest, P\'azm\'any P{\'e}ter s{\'e}t\'any 1/A, Hungary}

\author{A. Csord\'as}
\affiliation{Department of Physics of Complex Systems,E{\"o}tv{\"o}s University,H-1117 Budapest, P\'azm\'any P{\'e}ter s{\'e}t\'any 1/A, Hungary}
\affiliation{MTA-ELTE Theoretical Physics Research Group, P\'azm\'any P. S\'et\'any 1/A, H-1117 Budapest, Hungary}

\author{J. Cserti}
\affiliation{Department of Physics of Complex Systems,E{\"o}tv{\"o}s University,H-1117 Budapest, P\'azm\'any P{\'e}ter s{\'e}t\'any 1/A, Hungary}

%\wideabs{

\begin{abstract}
Silicene systems, due to the buckled structure of the lattice, manifest remarkable intrinsic spin-orbit interaction triggering a topological phase transition in the low-energy regime.
Thus, we found that protected edge states are present in silicene antidots and dots, being polarized in valley-spin pairs.
We have also studied the effect of the lattice termination on the properties of the single electron energy levels and electron density distribution of silicene antidots and dots situated in a perpendicular magnetic field.
Our calculations confirmed that the topological edge states are propagating over the perimeter of the antidot/dot for both ideal or realistic edge termination containing roughness on the atomic length scale.
The valley polarization and the slope of the energy line as a function of the magnetic field is, however, reduced when the antidot or dot has a rough edge.
\end{abstract}


\maketitle

\section{Introduction}

Silicene, a monolayer of silicon atoms forming a two-dimensional honeycomb lattice, has been recently successfully synthesized.\cite{silicene_synth1,silicene_synth2,silicene_synth3,silicene_synth4,silicene_synth5,silicene_synth6,silicene_synth7,silicene_synth8,silicene_synth9}
Several theoretical works indicate that the creation of a quasifreestanding silicene is also possible on particular substrates.\cite{freestanding1,freestanding2,freestanding3}
The low energy excitations of silicene can be described by two Dirac cones akin to graphene.
However, because of the large ionic radius of silicon, the two sublattices are arranged into two parallel planes.
Due to this buckled structure the six-fold symmetry typical for the graphene sheet is broken.
Consequently, silicene is expected\cite{soi1,soi2,quantumHall,soi3} to have a strong intrinsic spin-orbit interaction (SOI) compared to graphene\cite{kane_mele}.
According to first principle estimates, a topologically nontrivial band gap of the order of $\sim10$ meV\cite{soi3,freestanding1} is induced by the intrinsic SOI at the Dirac points.
In addition, the theoretical quantum Hall\cite{quantumHall} and quantum spin Hall\cite{quantumSpinHall} studies on silicene revealed a complex sequence of Landau levels\cite{LL_in_Ez} (LLs) and Hall plateaus,\cite{quantumHall} thus making the theoretical investigations of silicene nanostructures essential.
Recent theoretical studies predict the emergence of protected edge states (PES) in triangular silicene nanodiscs\cite{nanodisc} and in nanoribbons,\cite{nanoribbon,nanoribbon2} explained by the same mechanism as the quantum spin Hall effect in graphene.\cite{kane_mele}
In spite of such successful theoretical studies, the experimental realization of silicene based nanostructures is currently unavailable for the experimentalists.
In turn, the electric transport measurements on other two-dimensional topological insulator materials, including HgTe quantum wells\cite{HgTe1,HgTe_imaging,HgTe_ferromag} and InAs/GaSb heterostructures,\cite{InAs} have successfully demonstrated the role of the PESs in the charge\cite{HgTe_imaging,InAs} and spin transport processes.\cite{HgTe1,HgTe_ferromag}

According to these works, the low energy physics of quantum antidots (QAD) and quantum dots (QD) is also expected to be modified when a remarkable intrinsic SOI is present in the system.
The study of graphene QAD\cite{antidot,minAntidot,transport_antidot,GAL,DOSinGAL,transport,graphene_antidot_theo} and QD\cite{graphene_dot1,graphene_dot_meres,graphene_dot_meres2,graphene_dot_meres3,graphene_dot_theo1,graphene_dot_theo2} systems was addressed several times predicting possible experimental applications.
The physical properties of silicene based nanostructures are even more complex, since the SOI induced band gap can be controlled via several experimentally realistic methods.\cite{nanodisc,TI_inhomE,LL_in_Ez,photoradiation,antiferro}

In this paper we have studied the dependence of the single electron energy levels of circular silicene QADs and QDs on the edge termination.
To this end, we calculated the energy eigenvalues of QADs and QDs by the continuous model imposing the infinite mass boundary condition (IMBC) and compared them to the energy levels obtained within the tight binding framework imposing different boundary conditions.
We also studied the electron density distribution around the QADs and QDs for the different boundary conditions.
Similar studies were conducted on graphene QDs of triangular and hexagonal shape to explore the effect of the boundary condition on the energy eigenvalues of the QD.\cite{graphene_dot_theo1}
In our calculations, however, we examine the physical properties of QAD and QD systems considering imperfect edges containing irregularities on atomic length scale.

In addition, the intrinsic SOI in silicene systems introduces a topologically nontrivial insulator phase in the bulk band gap, giving rise to a PESs localized to the edge of the QAD or QD.
We will show that the PESs are also present in the systems when a perpendicular magnetic field is applied to the QADs or QDs.
In addition, these edge states are polarized into valley-spin pairs due to the intrinsic SOI.

The rest of the paper is organized as follows.
In Sec.~\ref{sec:models} we present the tight binding and continuous models used in our calculations.
In Sec.~\ref{sec:Gantidot} we study the energy eigenvalues and the electron density distribution in QADs neglecting the SOI. Here we present our numerical results obtained from the tight-binding calculations and compare them to the predictions of the continuous model.
In Sec.~\ref{sec:Gdot} we show that our findings on QADs are also valid to QDs.
In Sec.~\ref{sec:Siantidot} and \ref{sec:Sidot} we study the role of the SOI in the properties of the energy eigenvalues of silicene QADs and QDs, respectively.
Finally we summarize our work in Sec.~\ref{sec:summary}.

\section{Theoretical models} \label{sec:models}

\subsection{Tight binding (TB) model}
The freestanding ballistic silicene system is described by the following TB Hamiltonian:\cite{soi3,model2}
%\begin{equation}
\begin{align}
  H_{TB} = &-\sum\limits_{\langle ij\rangle\alpha}\gamma_{ij} c_{i\alpha}^{\dagger}c_{j\alpha} + 
{\rm i}\frac{\Delta_{SO}}{3\sqrt{3}}\sum\limits_{\langle\langle ij\rangle\rangle\alpha\beta}\nu_{ij}c_{i\alpha}^{\dagger}\sigma_{\alpha\beta}^zc_{j\beta} \\
&-{\rm i}\frac{2}{3}\lambda_R\sum\limits_{\langle\langle ij\rangle\rangle\alpha\beta}\mu_{i}c_{i\alpha}^{\dagger}\left(\boldsymbol{\sigma}\times\frac{\boldsymbol{d}_{ij}}{|\boldsymbol{d}_{ij}|}\right)_{\alpha\beta}^zc_{j\beta}, \label{eq:TB_Hamiltonian}
\end{align}
%\end{equation}
where the first term stands for the usual graphenelike nearest neighbor hopping Hamiltonian with $\gamma_{ij}\equiv\gamma$.
The second and the third terms are the effective intrinsic SOI and the intrinsic Rashba SOI, respectively.
Notations $\langle ij\rangle$ and $\langle\langle ij\rangle\rangle$ mean nearest neighbor and next-nearest neighbor sites in the lattice.
Finally $\boldsymbol{\sigma}=\left(\sigma_x,\sigma_y,\sigma_z\right)^T$ is the vector of the Pauli matrices, $\mu_i=\pm1$ for the A (B) sites and $\nu_{ij}=\frac{\mathbf{d}_i\times\mathbf{d}_j}{|\mathbf{d}_i\times\mathbf{d}_j|}$, with the two nearest bonds $\mathbf{d}_i$ and $\mathbf{d}_j$ connecting the next-nearest neighbors $\mathbf{d}_{ij}$.
In our calculations we used the effective parameters \mbox{$\gamma=1.6$ eV}, $\Delta_{SO}=3.9$ meV, and $\lambda_R=0.7$ meV obtained from first principles.\cite{soi3}
The magnetic field in the system was incorporated by means of the Peierls substitution.\cite{peierls}
The nearest neighbor hopping amplitudes, for example, can be given as
\begin{equation}
 \gamma_{ij} = \gamma\;{\rm exp}\left( \frac{2\pi{\rm i}}{\phi_0} \int\limits_{\mathbf{R}_j}^{\mathbf{R}_i} \mathbf{A(\mathbf{r})}{\rm d}\mathbf{r} \right),
\end{equation}
where $\phi_0 = h/e$ is the flux quantum, $\mathbf{R}_i$ points to the $i$th site of the lattice and the vector potential describing the magnetic field $\mathbf{B} = (0,0,B_z)^T$ was chosen in Landau gauge
\begin{equation}
 \mathbf{A}(\mathbf{r}) = \left[\left( \mathbf{r}\times\hat{\mathbf{a}}\right)_zB_z\right]\hat{\mathbf{a}}\;, \label{eq:vectorpot}
\end{equation}
where the $z$ direction is perpendicular to the plane of the silicene sheet and $\hat{\mathbf{a}}$ is an arbitrary unit vector in the $xy$ plane.
The distance between the nearest neighbor sites in silicene lattice, projected into the $xy$ plane, is \mbox{$d_0=2.23$ \AA}.

The TB Hamiltonians of the QADs studied in our work were constructed as a hole of radius $R$ located in a strip of finite length $L$ and width $W$.
One can identify the $\{ i\}$ set of sites to be removed from the lattice by a relation $|\mathbf{r}_i-\mathbf{r}_0|<= R - \xi_i$, where $\mathbf{r}_i$ is the position of the $i$th site and $\mathbf{r}_0=(W/2,L/2)$.
In order to include irregularities over the edge of the QAD, we introduce a random variable $0\leq\xi_i\leq\xi_{{\rm max}}$ (of uniform distribution), where $\xi_{max}\ll R$ represents an atomic length scale.
After solving the eigenproblem of the corresponding Hamiltonian, one needs to separate the edge states localized to the edges of the strip from the bound states localized to the QAD.
%If the dimensions of the strip are large compared to the decay length\cite{decaylength} of the edge states, the bound states of the QAD become separated from the edge states since the wave function of the bound states exponentially vanish with the distance measured from the QAD.
The energy eigenvalues of the bound states are robust to small variations in the dimensions of the strip which provides an efficient way to numerically separate bound states from the edge states.\cite{graphene_antidot_sajat}
In our calculations we shall consider QADs with three types of boundaries: (i) IMBC, (ii) smooth, and (iii) rough edge boundary conditions.
Instead of removing the sites inside the circle, one can also apply a high potential $V_0$ on these sites ($V_0$ is comparable to the TB hopping amplitude $\gamma$) with opposite signs on the two sublattices.
Then this staggered potential generates an effective mass term inside the circle disallowing the wave functions to enter there.
Since the applied $V_0$ potential is much greater than any other energy scale in the studied low energy limit, we denote this type of the boundary condition as the IMBC.
We also have studied QADs terminated by a circular edge including irregularities over the perimeter described by the parameter $\xi_i$.
We consider a QAD with smooth edge, when the irregularities over the perimeter are suppressed by $\xi_{{\rm max}}=0$, and for rough edged QAD $\xi_{{\rm max}}=d_0$. 
At the edge of the QADs the dangling bonds of the sites were not compensated.

The TB Hamiltonian of the studied QDs was constructed similarly as for the QADs.
In the case of the IMBC the staggered $V_0$ potential was applied outside the circular shaped area of ideal silicene, while the Hamiltonians of smooth and rough edged QDs were created as the counterpart of the QADs with smooth and rough edges, respectively.
Since our goal here is to reveal the main physical properties of individual QADs/QDs, we did not study their statistical properties upon the distribution of the irregularities over their lattice termination.\\[0.1cm]

\subsection{The continuous model}
The low energy effective Hamiltonian $H_K$ around the Dirac point $K$ can be derived as the long-wave approximation of the TB Hamiltonian (\ref{eq:TB_Hamiltonian}) and reads as
\begin{equation}
 H_K = \begin{pmatrix}
        h_{11} & v_F\hat{\pi}_+ \\
	v_F\hat{\pi}_- & -h_{11}
       \end{pmatrix} , \label{eq:contHamiltonian1}
\end{equation}
where $v_F\approx5.52\times10^5$ m/s is the Fermi velocity and $\hat{\pi}_{i} = \hat{p}_{i}+|e|A_{i}(\mathbf{r})$ (with $i=x,y$) are the operators of the canonical momentum and $\hat{\pi}_{\pm} = \hat{\pi}_x \pm{\rm i} \hat{\pi}_y$.
In our analytical calculations we used a vector potential given in the symmetric gauge $\mathbf{A}(\mathbf{r}) = \frac{1}{2}\mathbf{r}\times\mathbf{B}$.
The SOI is included in the matrix element $h_{11} = -\Delta_{SO}\sigma_z - \sqrt{3}d_0\lambda_R/\hbar\left(\hat{\pi}_y\sigma_x - \hat{\pi}_x\sigma_y\right)$.
Following calculations in Refs.~\onlinecite{similar2andor,andor} we introduce the $\hat{a}_+-=\frac{l_B\hat{\pi}_+}{\sqrt{2}\hbar}$ and $\hat{a}_{-}=\hat{a}_+^{\dagger}$ operators, where $l_B=\sqrt{\hbar/(|eB_z|)}$ is the magnetic length.
In polar coordinates the representation of these operators reads:
\begin{subequations} \label{eq:a_pm}
\begin{equation}
 \hat{a}_{-} = \frac{e^{-{\rm i}\varphi}}{\sqrt{2}{\rm i}}\left( \frac{\partial}{\partial\tau} - \frac{{\rm i}}{\tau}\frac{\partial}{\partial\varphi} + \frac{\tau}{2}\right)
\end{equation}
and
\begin{equation}
 \hat{a}_{+}= \frac{e^{{\rm i}\varphi}}{\sqrt{2}{\rm i}}\left( \frac{\partial}{\partial\tau} + \frac{{\rm i}}{\tau}\frac{\partial}{\partial\varphi} - \frac{\tau}{2}\right),
\end{equation}
\end{subequations}
with the dimensionless radial coordinate $\tau=r/l_B$.
Then the Hamiltonian in Eq.~(\ref{eq:contHamiltonian1}) becomes
\begin{equation}
 H_K = \begin{pmatrix}
        h_{11} & \hbar\omega_c \hat{a}_+ \\
	\hbar\omega_c \hat{a}_- & -h_{11}
       \end{pmatrix}, \label{eq:contHamiltonian2}
\end{equation}
with 
\begin{equation}
 h_{11}=-\Delta_{SO}\sigma_z - \sqrt{\frac{3}{2}}\frac{d_0\lambda_R}{2l_B}{\rm i}\left(\hat{a}_-\sigma_+ - \hat{a}_+\sigma_-\right) \label{eq:h11_witha}
\end{equation}
and $\hbar\omega_c = \sqrt{2v_F^2\hbar|eB|}$ is the cyclotron frequency.
In addition, we introduce a basis of square-integrable functions to find the eigenstates of Hamiltonian (\ref{eq:contHamiltonian2}).
Following the reasoning of Refs. \onlinecite{similar2andor,andor}, we aimed at choosing a basis set that would establish one-to-one correspondence between the base functions when the operators $a_{\pm}$ are applied on them.
[The action of the operators $a_{\pm}$ on a real-space function is fully determined by their differential representation given in Eq.(\ref{eq:a_pm}).]
Moreover, for QADs of radius $R$ (so $r\geq R$) the basis functions must vanish as $r\rightarrow\infty$ (we assume that the QAD is located far away from the edges of the sample), while in the case of QDs of radius $R$ ($0\leq r\leq R$) the basis functions need to be regular at the center of the QD.
Thus, the appropriate base functions can be written as:
\begin{equation}
 f_{a,m}(\delta,\varphi) = e^{{\rm i}m\varphi}\delta^{|m|/2}e^{-\delta/2}\;U(a,|m|+1,\delta), \label{eq:basis_antidot}
\end{equation}
where $\delta=\tau^2/2$ and the integer $m$ is the angular momentum quantum number.
For QAD the function $U$ is the Kummer's function\cite{abramovitz} with real parameter $a$ and being square-integrable in the range $\delta\in[\delta_0,\infty]$, where 
$\delta_0=\frac{(R/l_B)^2}{2}$ is the missing magnetic flux in units of $\phi_0$ inside the QAD.
Since the Kummer's function $U$ is divergent as $r\rightarrow0$ (except for parameter $a$ being a nonpositive integer), in case of QDs one needs to use the other Kummer's function in Eq.~(\ref{eq:basis_antidot}), which is regular for $r\rightarrow0$ and denoted by $M$ in this work.\cite{abramovitz}
%Conversely, the functions $M$ are not suitable to describe the electron states of QAD systems studied in this work either, since $M$ does not vanish for $r\rightarrow\infty$.

In the following, we limit our calculations in energy to the vicinity of the Dirac points.
Since the strength of the Rashba SOI is proportional to the momentum measured from the Dirac point, for simplicity, we neglect the Rashba SOI in our further calculations.
Thus the Hamiltonian in Eq.~(\ref{eq:contHamiltonian2}) becomes block diagonal in the spin space $s=\pm1$, corresponding for the up-standing ($\uparrow$) and down-standing ($\downarrow$) spin, respectively.
Later on, in the framework of the TB model, we check that the presence of Rashba SOI does not change our results significantly.

To solve the eigenproblem of the two block-diagonal parts of the Hamiltonian $H_K$ for the bulk system we use the following wave function ansatz
\begin{subequations} \label{eq:wavefunc}
\begin{equation}
 \Psi_{a,m,s} = \begin{pmatrix}
                \Lambda_{a,m,s}f_{a-1,m+1} \\
		f_{a,m}
               \end{pmatrix} \quad \textrm{if } m<0, \label{eq:wavefunc_mneg}
\end{equation}
and
\begin{equation}
 \Psi_{a,m,s} = \begin{pmatrix}
                \Lambda_{a,m,s}f_{a,m+1} \\
		f_{a,m}
               \end{pmatrix} \quad \textrm{if } m\geq0.
\end{equation}
\end{subequations}
The energy eigenvalues $E_{a,m,s}$ and the $\Lambda_{a,m,s}$ amplitudes can be obtained by straightforward calculations. The details of the calculations are presented in the Appendix.
From an eigenproblem point of view we omit the normalization of the wave functions to unity in Eq.~(\ref{eq:wavefunc}).
However, during the evaluation of other physical quantities, including the electron density distribution, the normalization factor has been included.
For QADs and QDs we obtain the same energy eigenvalues, namely for $m<0$
\begin{subequations}
\begin{equation}
 E^{\pm}_{a,m,s} = \pm\sqrt{\left(1-a\right)(\hbar\omega_c)^2 + \Delta_{SO}^2}, \label{eq:E_B}
\end{equation}
while for $m\geq0$
\begin{equation}
 E^{\pm}_{a,m,s} = \pm\sqrt{\left(m+1-a\right)(\hbar\omega_c)^2 + \Delta_{SO}^2}. \label{eq:E_A}
\end{equation} 
\end{subequations}
When the radius of the QAD is zero, we recover the LLs of the bulk silicene system.\cite{quantumHall,LL_in_Ez}
For QAD the basis functions (\ref{eq:basis_antidot}) are regular at $r\rightarrow0$ when $a=1-n$, and $n$ is a non-negative [positive] integer for $m<0$ [$m\geq0$].
However, for $n=0$ and \mbox{$m<0$} the parameter $\Lambda_{a,m,s}$ in Eq.(\ref{eq:wavefunc_mneg}) is formally divergent (for details see the Appendix).
In this case the wave function must be regularized by a normalization factor.
The energy levels of the LLs can be then written as $E_{LL}^{\pm}(n,m,s)=E^{\pm}_{1-n,m,s}$ in agreement with the literature.\cite{quantumHall,LL_in_Ez}
Notice that the LLs depend on the non-negative values of the angular momentum quantum number $m$.
For QD of infinite radius one can recover identical results for the LLs.

Now we turn to the study of QADs/QDs of finite radius.
For simplicity, we impose the IMBC over the edge of the QAD/QD, namely $\Psi(1)/\Psi(2)\equiv\xi{\rm i}e^{{\rm i}\varphi}$, where $\Psi(1)$ and $\Psi(2)$ denote the upper and lower component of the wave function (\ref{eq:wavefunc}) at the edge of the QAD/QD, respectively.
The parameter $\xi=1$ [$\xi=-1$] corresponds to the $K$ [$K'$] valley.
Solving the secular equation one can calculate the $E^{\pm}(\xi,n,m,s)$ energy eigenvalues for the QAD and QD systems, where $n$ is the radial quantum number.
%The electron-hole symmetry assures the $E^{\pm}_{X}(\xi,n,m)=-E^{\mp}_{X}(-\xi,n,m)$ relation between the energy eigenvalues related to the different valleys labeled by $\xi$.
For QAD, inserting the wave function ansatz given in Eq.~(\ref{eq:wavefunc}) into the secular equation, we obtain
\begin{equation}
 \xi\frac{E+s\Delta_{SO}}{\sqrt{\delta}\hbar\omega_c} = \frac{U(a,m+2,\delta_0)}{U(a,m+1,\delta_0)}, \label{eq:secular_antidot}
\end{equation}
while in the case of a QD for $m<0$
\begin{subequations} \label{eq:secular_dot}
\begin{equation}
 \xi\sqrt{\delta}\frac{E+s\Delta_{SO}}{m\hbar\omega_c} = \frac{M(a-1,|m|,\delta_0)}{M(a,|m|+1,\delta_0)}
\end{equation}
and for $m\geq0$
\begin{equation}
 \sqrt{\delta}\frac{E-s\Delta_{SO}}{\xi(m+1)\hbar\omega_c} = \frac{M(a,m+1,\delta_0)}{M(a,m+2,\delta_0)}.
\end{equation}
\end{subequations}
For QAD the parameter $a=a(E,m)$ is expressed from Eq.~(\ref{eq:E_A}) for both negative and positive $m$.
For QD, on the other hand, the parameter $a=a(E,m)$ is expressed from Eqs.~(\ref{eq:E_B}) and (\ref{eq:E_A}) for $m<0$ and $m\geq0$, respectively.
It can be also noted that the secular equations of the QAD for different signs of the angular momentum $m$ can be unified due to the Kummer transformation\cite{abramovitz}, namely that $U(a,b,z)=z^{1-b}U(1+a-b,2-b,z)$. 

Due to the electron-hole symmetry characteristic for our models, we will study only the positive energy regime in our calculations.
In the following sections we compare the energy eigenvalues calculated by the continuous model to the energy eigenvalues obtained in the TB framework imposing different 
boundary conditions and examine which features of the energy eigenvalues are manifested by the continuous model.

\section{QAD without the spin-orbit interaction (graphene QAD)} \label{sec:Gantidot}

In this section we focus on the energy eigenvalues and electron density distribution of a QAD.
First we set the strength of the SOI to zero.
By disregarding the SOI terms in Hamiltonian (\ref{eq:TB_Hamiltonian}) one ends up with Hamiltonian typical also for graphene systems.
The graphene QAD has already been studied in Ref.~\onlinecite{antidot}, hence our results partially overlap with this work.
Without the SOI the energy eigenvalues of graphene systems are degenerated in the spin degree of freedom. 
In our calculations for graphene we used the TB parameters \mbox{$\gamma=3$ eV} and \mbox{$d_0=1.42$ \AA}.
\begin{figure}[thb]
\includegraphics[scale=0.5]{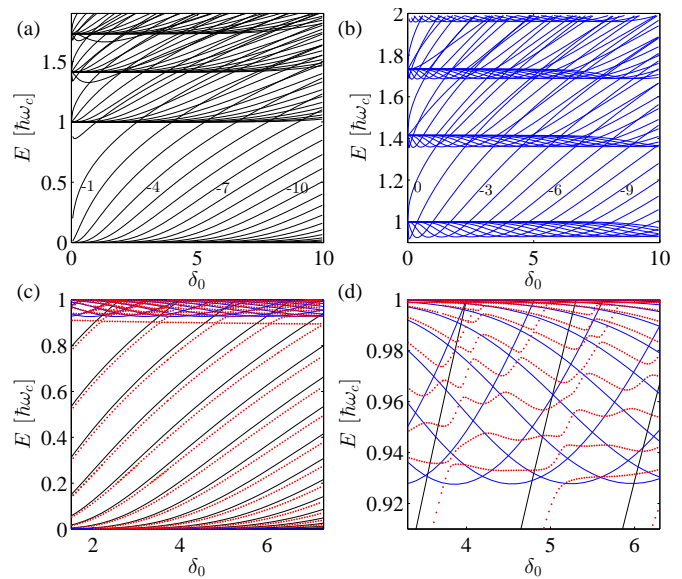}
\caption{(Color online) (a) and (b) Energy eigenvalues of a graphene QAD as a function of $\delta_0$ calculated by Eq.~(\ref{eq:secular_antidot}) for $m=-20,\dots,3$ at valleys $K$ and $K'$, respectively.
The numbers in the figures indicate the $m$ dependency of the energy eigenvalues, starting with $m=n-1$ above [below] the $n$th [($n+1$)th] LL.
(c) Energy eigenvalues calculated by TB model (red dotted line) imposing the IMBC (see the text for details) are in good agreement with the results of the continuous model calculated at $K$ (black line) and $K'$ (blue line) valleys.
(d) For better view, the data of figure (c) are enlarged around the energy eigenvalues corresponding to states of the valley $K'$. 
The degeneracies at the crossings of the lines corresponding to the continuous model are lifted by the TB model.
(The TB calculations were performed for QAD of radius $R=11.1$ nm)
\label{fig:graphene_antidot}} 
%%\vspace{-5mm}
\end{figure}
In Fig.~\ref{fig:graphene_antidot} we show the energy eigenvalues calculated by the continuous model [see Figs.~\ref{fig:graphene_antidot}(a) and \ref{fig:graphene_antidot}(b)] and compare them 
to the TB model [see Figs.~\ref{fig:graphene_antidot}(c) and \ref{fig:graphene_antidot}(d)] imposing 
the IMBC in both models.
One can see a good agreement between the results of the continuous and TB model.
In the TB framework, however, one can observe the presence of highly localized states, which is a characteristic feature in TB models.\cite{localized_states}
In particular, Fig.~\ref{fig:graphene_antidot}(c) shows TB energy eigenvalues of states localized only to one side of the QAD close to energy $\sim0.9\hbar\omega_c$.
The energy line of these states have a weak dependence on the magnetic field and are missing in the results of the continuous model, since they carry zero current around the QAD. 
At low magnetic fields, when the cyclotron radius becomes larger than the radius of the QAD, the energy line of the states tend to the closest LL.
Our results are in good agreement with other theoretical studies in the literature.\cite{antidot,graphene_antidot_nothole,graphene_antidot_sajat}
The differences between the results of the continuous and TB model can be explained by the mixing of electron states of different 
angular momentum $m$ and to the scattering processes between the valleys [see the avoided crossings around the black lines in Fig.~\ref{fig:graphene_antidot}.(d)].
The lattice structure of the TB model breaks the rotation symmetry of the continuous model, hence the angular momentum $m$ is no longer a good quantum
number in the TB model.
For the IMBC, however, the mixing of the states in the TB wave function is low even in QADs of moderate radius.
\begin{figure}[thb]
\includegraphics[scale=0.43]{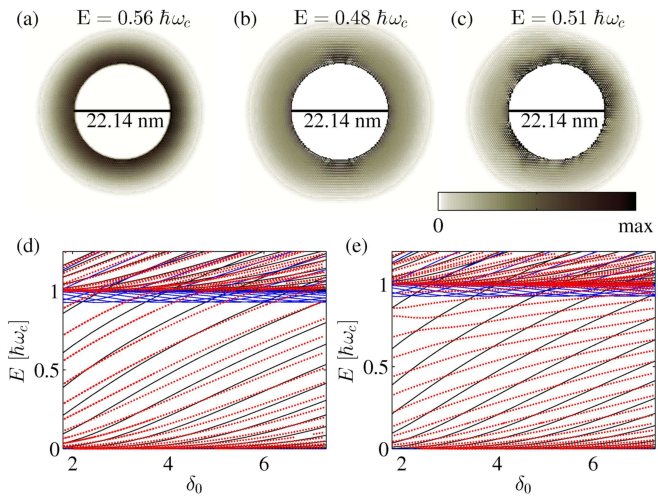}
\caption{(Color online) The electron density distribution of the QAD for (a) IMBC, (b) smooth edge, and (c) rough edge boundary condition at $\delta_0=3.72$.
The bright (dark) colors corresponds to low (high) density areas.
(d) [(e)] Energy eigenvalues of a graphene QAD as a function of  $\delta_0$  calculated by the TB model (red dotted line) imposing a smooth [rough] boundary condition (see the text for details) compared to the results of the continuous model calculated at $K$ (black line) and $K'$ (blue line) valleys. 
(The TB calculations were made for QAD of radius $R=11.1$ nm.)
\label{fig:graphene_antidot2}} 
%%\vspace{-5mm}
\end{figure}
Indeed, as one can see in Fig.~\ref{fig:graphene_antidot2}(a), the calculated electron density distribution $\rho(\mathbf{r}) = \sum\limits_{s=\pm1}|\Psi_s(\mathbf{r})|^2$ is isotropic for the IMBC.
If the QAD is terminated by an edge, like the case plotted in Figs.~\ref{fig:graphene_antidot2}(b) and \ref{fig:graphene_antidot2}(c), the electron density becomes anisotropic.
In order to study the effect of the lattice termination on the electron density distribution,
we have used the TB model to create QADs terminated by circular edge including irregularities over the perimeter.
For smooth edge the irregularities were suppressed and the anisotropy in electron density distribution is weak [see Fig.~\ref{fig:graphene_antidot2}(b)], 
while for rough edge including pronounced irregularities the anisotropy is significant [see Fig.~\ref{fig:graphene_antidot2}.(c)].
Also, as the edge of the QAD becomes rough, the mixing of the valleys in the wave functions is more and more pronounced resulting in a short-distance oscillations in the electron density distribution, that is shown in Figs.~\ref{fig:graphene_antidot2}(b) and \ref{fig:graphene_antidot2}(c).
Like electron density distribution, the energy eigenvalues are also affected by the properties of the edge termination.
Figures \ref{fig:graphene_antidot2}(d) and \ref{fig:graphene_antidot2}(e) show the calculated TB energy eigenvalues of a QAD with a smooth and a rough edge, respectively.
The slope of the energy line becomes smaller for the QAD with rough edge, while the energy line corresponding to QAD with a smooth edge
 are close to the results of the continuous model.
 Hence, if the length scale of the QAD is much larger than the lattice constant, so the irregularities over the perimeter can be neglected, one can expect 
 that the physical properties of the QAD would tend to the properties of a QAD enclosed by the IMBC.

\section{QD without the spin-orbit interaction (graphene QD)} \label{sec:Gdot}

We now turn to the studies of QD systems.
We show that the conclusions made for QADs in the previous section are also relevant for QD systems.
\begin{figure}[thb]
\includegraphics[scale=0.5]{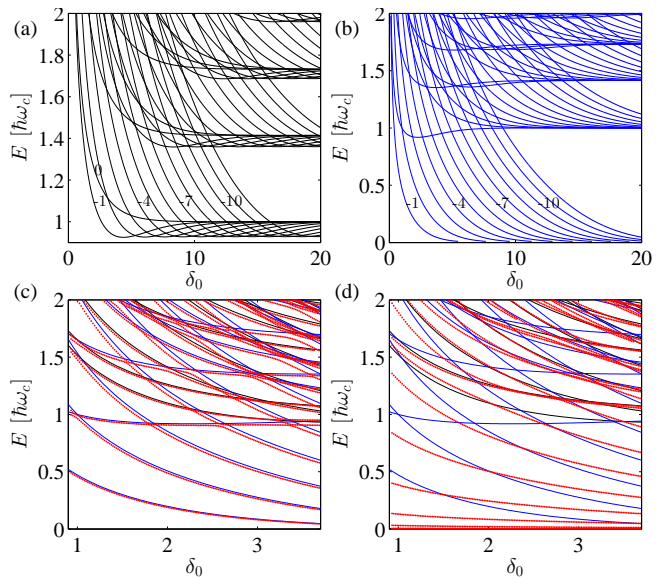}
\caption{(Color online) (a) and (b) Energy eigenvalues as a function of the magnetic flux inside the graphene QD calculated by Eq.~(\ref{eq:secular_dot}) for $m=-12,\dots,3$ at valleys $K$ and $K'$, respectively. 
The numbers in the figures indicate the $m$ dependency of the energy eigenvalues, starting with $m=n-1$ above [below] the $n$th [($n+1$)th] LL.
(c) [(d)] Energy eigenvalues calculated by TB model (red dotted lines) imposing the infinite mass [rough edge] boundary condition compared to the results of the continuous model calculated at $K$ (black line) and $K'$ (blue line) valleys. 
The TB energy eigenvalues calculated by the IMBC are in good agreement with the results of the continuous model.
(The TB calculations were made for QD of radius $R=11.7$ nm.)
\label{fig:graphene_dot}} 
%%\vspace{-5mm}
\end{figure}
In the following calculations we also turned off the SOI in our models, that is corresponding to the case of graphene QD.\cite{graphene_dot1,graphene_dot_meres}
Figures \ref{fig:graphene_dot}(a) and \ref{fig:graphene_dot}(b) show the energy eigenvalues of a graphene QD calculated by the continuous model imposing the IMBC at valleys $K$ and $K'$, respectively.
(The energy eigenvalues are degenerated in the spin degree of freedom.)
Our results are in good agreement with other theoretical studies in the literature.\cite{graphene_dot1,graphene_dot_meres}
One can see that in contrast to the case of QADs, the energy line tend to the energies of the LLs for large magnetic field.
Notice that in large magnetic field the cyclotron radius becomes small compared to the radius of the QD.
We also calculated the TB energy eigenvalues of QDs enclosed by the IMBC [see Fig.~\ref{fig:graphene_dot}(c)] and by a smooth lattice termination [see Fig.~\ref{fig:graphene_dot}(d)].
As in the case of graphene QAD, the energy eigenvalues calculated within the continuous model by Eq.~(\ref{eq:secular_dot}) 
are in good agreement with the TB energy eigenvalues obtained by the IMBC.
However, one can observe avoided crossings of the TB energy line due to the mixing of states with different angular momentum and due to the valley-valley 
scattering processes induced by the short range changes of the mass term around the perimeter of the QD.
In the case of a QD with smooth edge termination, the slope of the TB energy line is, similarly to QADs, smaller than the slope of the energy line
 calculated by the continuous model [see Fig.~\ref{fig:graphene_dot}(d)].
We also checked that the differences between the results of the continuous and TB model further increases if the edge of the QD is rough.
Thus, in realistic experiments where the edge termination of the prepared QD always includes irregularities, one expects that the slope of the single electron energy line is smaller than the one predicted by the continuous model imposing the IMBC.
Indeed, in Ref.~\onlinecite{graphene_dot_meres} the experimentally obtained slope of the single electron energy line was about three times lower than in the theoretical model imposing the IMBC.\cite{graphene_dot1}

The calculated electron density distribution (not presented here) shows similar behavior as in the case of QAD.
For IMBC the electron density distribution is isotropic, while for QDs with smooth and rough edge the electron density distribution becomes anisotropic.

\section{QAD including the spin-orbit interaction} \label{sec:Siantidot}

After examining the effect of the edge properties on the energy eigenvalues of QADs and QDs with zero SOI, we now examine how the intrinsic SOI changes the results of the previous sections.
The intrinsic SOI lifts the spin degeneracy of the eigenvalues typical to graphene structures.
In addition the intrinsic SOI introduces topological phase transition in the low energy regime $-\Delta_{SO}\leq E\leq\Delta_{SO}$, leading to the emergence of PES among the magnetic bound states.
\begin{figure}[thb]
\includegraphics[scale=0.5]{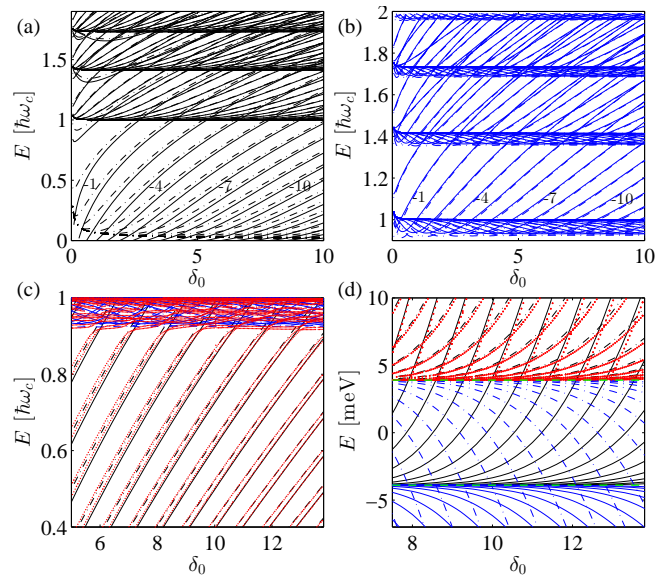}
\caption{(Color online) (a) and (b) Energy eigenvalues as a function of the missing flux inside the silicene QAD for $\uparrow$ (solid line) and $\downarrow$ (dash-dotted line) spin states calculated by Eq.~(\ref{eq:secular_antidot}) for $m=-30,\dots,3$ at valleys $K$ and $K'$, respectively.
The numbers in the figures indicate the $m$ dependency of the energy eigenvalues, starting with $m=n-1$ above [below] the $n$th [($n+1$)th] LL.
(c) Energy eigenvalues calculated by TB model (red dotted line) imposing the IMBC (see the text for details) are in good 
agreement with the results of the continuous model calculated at $K$ (black line) and $K'$ (blue line) valleys.
(d) The enlarged plot of (a) in the low energy regime and compared to the results of the TB model.
The lines corresponding to $\uparrow$ [$\downarrow$] spin in both valleys tend to the lowest LL of energy $\Delta_{SO}$ [$-\Delta_{SO}$].
For a better view the TB results are displayed only in the energy range $E>\Delta_{SO}$.
The dashed green lines indicates the energy of the lowest LL.
(The TB calculations were made for QAD of radius $R=17.4$ nm and with $\Delta_{SO}=3.9$ meV.)
\label{fig:silicene_antidot}} 
%%\vspace{-5mm}
\end{figure}
Figures \ref{fig:silicene_antidot}(a) and \ref{fig:silicene_antidot}(b) show the calculated energy eigenvalues of a silicene QAD corresponding to valleys $K$ and $K'$, respectively.
The energy eigenvalues were calculated within the continuous model using Eq.~(\ref{eq:secular_antidot}).
The role of the intrinsic SOI is significant in the low energy regime, the energy line corresponding to $\downarrow$ [$\uparrow$] spin tend to the lowest LL of energy $\Delta_{SO}$ [$-\Delta_{SO}$] for both valleys.
%Csillus a mokus
The energy eigenvalues inside the bulk band gap $[-\Delta_{SO},\Delta_{SO}]$ are originating from the meeting of two topologically different phases\cite{top_states,top_states2}.
Inside the circle, due to the IMBC, the effective mass has the same sign in both valleys.
(Notice that in the TB model the IMBC was modeled by a high potential of opposite signs on the two sublattices.)
On the other hand, outside the circle the effective mass, due to the intrinsic SOI, has a different sign for the two spin/valley states.
%Since the effective mass changes the sign if the valley or the spin is changed, one can conclude that the effective mass has the same sign for states corresponding to different valleys and opposite spins.
Particularly, the PESs shown in Fig.~\ref{fig:silicene_antidot}.(d) and crossing the bulk band gap stand for for the $(K,\uparrow)$ and $(K',\downarrow)$ pairs
[the pair $(K,\uparrow)$ corresponds to a state in the valley $K$ with $\uparrow$ spin], 
while the energy line corresponding to ordinary quantum Hall states [pairs $(K,\downarrow)$ and $(K',\uparrow)$] do not cross the bulk band gap, since for these states the effective mass has the same sign as in the inside of the circle.
The nature of the PESs differs from the quantum Hall states localized to the QAD, since the PESs are also present in the system and remain localized to the edge, when the magnetic field vanishes and the quantum Hall states evolve into delocalized states.
Another robust feature of the PES is that the current of the different valley-spin polarized pairs shown in Fig.~\ref{fig:silicene_antidot}.(d) is flowing in the opposite direction to each other.
(Notice that the current whirling around the QAD is proportional to the derivate $-\frac{\partial E}{\partial\delta_0}$.\cite{persistent_current})

\begin{figure}[thb]
\includegraphics[scale=0.5]{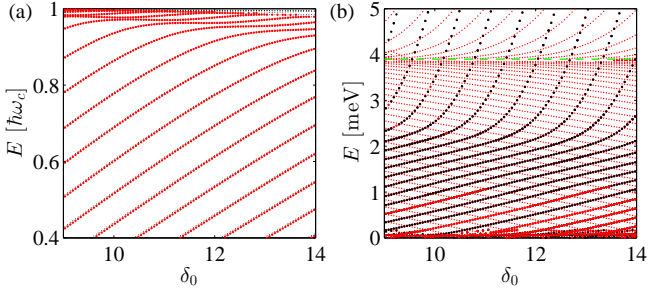}
\caption{(Color online) (a) Energy eigenvalues as a function of the missing flux inside the silicene QAD calculated from the TB model (red dotted line) imposing the smooth edge boundary condition.  Filled black (opened red) circles correspond to $\uparrow$ ($\downarrow$) spin states.
(b) The enlarged plot in the low energy regime. In the bulk band gap of $[-\Delta_{SO},\Delta_{SO}]$ the silicene QAD exhibits PESs that are induced by the intrinsic SOI.
The dashed green line at $E=\Delta_{SO}$ indicates the energy of the lowest LL.
(The TB calculations were made for QAD of radius $R=11.6$ nm and with $\Delta_{SO}=3.9$ meV.)
\label{fig:silicene_antidot2}} 
%%\vspace{-5mm}
\end{figure}

Similarly to graphene QAD, the slope of the energy line calculated for smooth edge lattice termination was lower than the slope of the energy line calculated with the IMBC [see Fig.~\ref{fig:silicene_antidot2}(a)].
The slope of the energy line of the PESs in the low energy regime is even more reduced [see Fig.~\ref{fig:silicene_antidot2}(b)].
The physical properties of these states are similar to the PES analyzed in the previous paragraph, however, in this case the mixing of the valleys in the electron states is stronger due to valley-valley scattering processes at the edge termination.
We also checked that the presence of a small Rashba SOI term in the TB model mixes the spin up and spin down states and induces avoided crossings between the energy line in Fig.~\ref{fig:silicene_antidot2}(b).
The energy gap of the avoided crossings can be estimated to $\sim\frac{d_0}{l_B}\lambda_R$ by the coupling strength of the Rashba-type SOI given in Eq.~(\ref{eq:h11_witha}).
For $B=1$T the induced gap is \mbox{$\sim 3.5\times10^{-3}\lambda_R$} that is an order of $\mu$eV for the $\lambda_R$ coupling strength given in Sec.~\ref{sec:models}.
Thus the effect of the Rashba SOI on the obtained results is marginal.

For QAD with a rough edge, the slope of the obtained energy line was even more smaller.
Similarly to the graphene QAD, the anisotropy of the electron density distribution and the slope of the energy line are closely connected properties.
For rough edged QAD, when the electron density distribution becomes anisotropic, the slope of the energy line is simultaneously reduced.

\section{QD including the spin-orbit interaction} \label{sec:Sidot}

\begin{figure}[thb]
\includegraphics[scale=0.5]{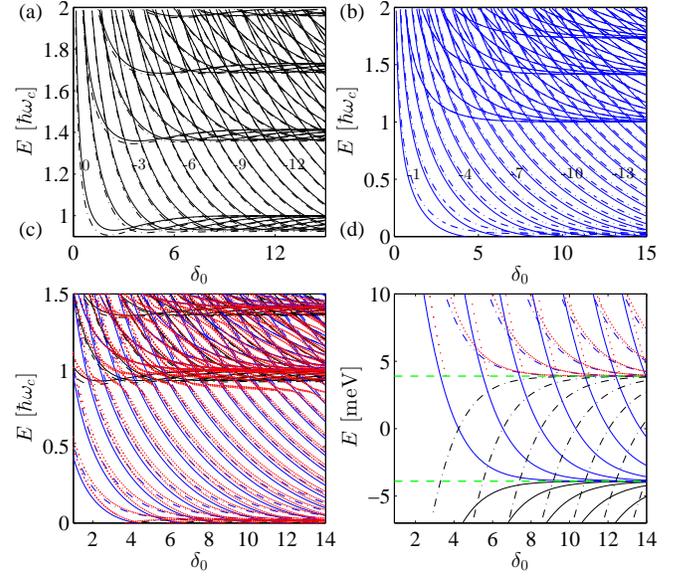}
\caption{(Color online) (a) and (b) Energy eigenvalues as a function of the magnetic flux inside the silicene QD for $\uparrow$ (solid line and filled circles) and $\downarrow$ (dash-dotted line and opened circles) spin states calculated by Eq.~(\ref{eq:secular_dot}) for $m=-12,\dots,3$ at valleys $K$ 
and $K'$, respectively. 
The numbers in the figures indicate the $m$ dependency of the energy eigenvalues, starting with $m=n-1$ above [below] the $n$th [($n+1$)th] LL.
(c) Energy eigenvalues calculated by TB model (red dotted line) imposing the IMBC (see the text for details) are in good 
agreement with the results of the continuous model calculated at $K$ (black line) and $K'$ (blue line) valleys.
(d) The enlarged plot shows the low energy regime. The lines corresponding to $\uparrow$ [$\downarrow$] spin in both valleys tend to the lowest LL of energy $\Delta_{SO}$ [$-\Delta_{SO}$].
For a better view the TB results are displayed only in the energy range $E>\Delta_{SO}$.
The dashed green lines indicates the energy of the lowest LL.
(The TB calculations were made for QD of radius $R=17.4$ nm and with $\Delta_{SO}=3.9$ meV.)
\label{fig:silicene_dot}} 
%%\vspace{-5mm}
\end{figure}

Finally, we also examine the role of the intrinsic SOI in the energy eigenvalues of silicene QDs confined by the boundary conditions described in Sec.~\ref{sec:models}.
Though it has been demonstrated theoretically for graphene QDs that a charge confinement can also be created by a spatial modulation of the mass term of the Dirac quasiparticles.\cite{graphene_dot_theo2}
In silicene, a nonhomogeneous mass term could be induced, for example, by controlling the strength of the intrinsic SOI.
The intrinsic SOI, similarly to the case of silicene QAD, also lifts the spin degeneracy of the energy eigenvalues.
Figures \ref{fig:silicene_dot}(a) and \ref{fig:silicene_dot}(b) show the calculated energy eigenvalues of a silicene QD corresponding to valleys $K$ and $K'$, respectively.
The energy eigenvalues were calculated within the continuous model using Eq.~(\ref{eq:secular_dot}).
Comparing the results of the continuous model to the TB energy eigenvalues calculated with the IMBC, we find again a good agreement between the two models as can be seen in Fig.~\ref{fig:silicene_dot}(c).
In the low energy regime the energy eigenvalues show the same peculiar behavior as in the case of silicene QAD, namely the lines corresponding to $\uparrow$ states tend to the $n=0$ LL of energy $\Delta_{SO}$, while the energy line of $\downarrow$ states tend to $-\Delta_{SO}$ [see Fig.~\ref{fig:silicene_dot}(d)]. 
In the energy range of the bulk band gap the PESs are polarized in valley-spin pairs and propagate in the opposite direction. 
\\
\begin{figure}[thb]
\includegraphics[scale=0.5]{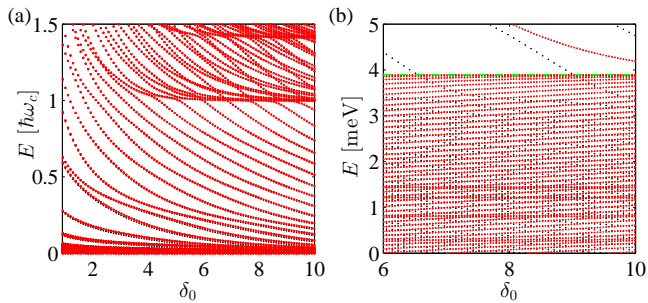}
\caption{(Color online) (a) Energy eigenvalues as a function of the magnetic flux inside the silicene QD calculated from the TB model (red dotted line) imposing a smooth edge boundary condition.
Filled black (empty red) circles correspond to $\uparrow$ ($\downarrow$) spin states.
(b) The enlarged plot shows the low energy regime. In the bulk band gap of $[-\Delta_{SO},\Delta_{SO}]$ the silicene QD exhibits PES due to the presence of the intrinsic SOI in the system.
%The red solid (black dotted) line correspond to $\uparrow$ ($\downarrow$) spin states.
The dashed green  line at $E=\Delta_{SO}$ indicates the energy of the lowest LL.
(The TB calculations were made for QD of radius $R=17.4$ nm and with $\Delta_{SO}=3.9$ meV.)
\label{fig:silicene_dot2}} 
%%\vspace{-5mm}
\end{figure}
Similarly to graphene QD, the slope of the TB energy line calculated for QD with a smooth edge is smaller than the slope of the lines obtained by models with the IMBC [see Fig.~\ref{fig:silicene_dot2}(a)].
However, the qualitative properties of the eigenstates are similar to QDs with the IMBC.
In the energy range of the bulk band gap shown in Fig.~\ref{fig:silicene_dot2}(b) one can observe the energy eigenvalues of the PES, that propagates in the opposite direction for opposite spin polarization.
The PESs, like for silicene QAD, are present in the system even when the magnetic field is zero.\cite{nanodisc} 
In addition, we numerically checked that the presence of a finite Rashba SOI does not change the obtained result significantly.

\section{Summary}
\label{sec:summary}

In this paper we showed that the boundary condition at the edge of QADs and QDs in materials possessing a hexagonal lattice structure have a great influence on their energy levels. 
Although, the lattice structure in the TB model breaks the rotational, the resulted mixing of the electron states of different angular momenta remains low.
We found that the influence of the boundary condition on the energy levels is rather significant.
For smooth edges the slope of the energy line as a function of the perpendicular magnetic field becomes lower than in systems with an ideal edge modeled by the IMBC.
In addition, the slope of the energy line of QADs and QDs with a rough edge is even smaller.
Our findings are further supported by the studies made on the electron density distribution.
Indeed, electron density distribution is anisotropic in rough edged systems and becomes isotropic as the edge of the QAD or QD becomes smoother.

As far as we know, the silicene based nanostructures are still unavailable for experimental purpose. However, we expect that typical Coulomb blockade experiments might provide a possible way to probe the excited states in these systems, as was demonstrated in graphene based nanostructures.\cite{graphene_dot_meres,graphene_dot_meres2,graphene_dot_meres3} 
Moreover, these experiments also showed that the traces of the excited states can also be observed by exploring the charge stability diagrams.\cite{graphene_dot_meres,graphene_dot_meres2,graphene_dot_meres3} 
In these measurements the typical size of the studied graphene sample was \mbox{$70-140$ nm} with a number of charge carriers in the system estimated to be an order of 10.
Also, the range of the magnetic flux used in our calculations can be converted to a magnetic field of strength $1-8$ T (assuming samples of characteristic size of \mbox{$100$ nm}), making our results relevant for experimental applications.
%Resonant transport tunneling via double QDs reported on GaAs/AlGaAs heterostructures seems to be also a promising experimental method for an accurate determination of the intrinsic lifetime and energy levels of discrete energy states in QDs.\cite{double_dot}

We expect that in realistic experiments on silicene the observed physical properties would be closer to the predictions of theoretical models with smooth or rough edge termination, due to the irregular shape of the QADs and QDs.
In the studied QADs we have also found a signature of highly localized edge states carrying zero current around the QAD.
%The emergence of these state is typical for hexagonal lattices with zigzag edges.
%The energy of these states has a low dependence on the magnetic field.
The presence of the intrinsic SOI in silicene systems lifts the spin degeneracy of the energy levels.
Moreover, in the bulk band gap induced by the intrinsic SOI we have identified PES generated on the interface of topologically different phases.
The PESs are well described by valley-spin polarized pairs due to the effective mass term induced by the intrinsic SOI and having different sign for the two valleys/spins.
This simple physical picture is, however, blurred by the presence of a finite Rashba type SOI and by the valley-valley scattering mechanism.
In experiments, the mixing of the valleys can be suppressed by a smooth edge tending to the ideal IMBC.
In silicene the small strength of the Rashba SOI close to the Dirac points also favors the emergence of the valley-spin polarized PESs.
We believe that the spin and valley polarized transport via the PESs can be utilized in various experimental applications.
A spin polarized current can be injected into the system by using, for example, ferromagnet contacts as it was demonstrated for HgTe quantum well structures.\cite{HgTe_ferromag}
In addition, the coupling of the valley and spin degrees of freedom is expected to provide further possibilities for the realization of valley engineering in silicene based nanostructures.
We believe that our conclusions are of general importance, and they are applicable for other confined ballistic silicene systems as well.

\section*{Acknowledgments}

We would like to acknowledge the support of the Hungarian Scientific Research Funds (OTKA) K108676 and K81492.

\appendix

\section*{Eigenproblem of the effective Hamiltonian $H_K$} \label{subsec:eigen}

\emph{Quantum antidot:}
In order to solve the eigenproblem of the Hamiltonian (\ref{eq:contHamiltonian2}) for QAD system, we first calculate the action of the operators $\hat{a}_{\pm}$ on the basis functions (\ref{eq:basis_antidot}):
\begin{subequations}
\begin{equation}
 \hat{a}_- f_{a,m} = \left\{\begin{matrix}
                                  {\rm i}a f_{a+1,m-1}  & \textrm{if } m\leq0 \\
				  {\rm i}(a-m) f_{a,m-1}  & \textrm{if } m>0
                                 \end{matrix}  \right.,
\end{equation}
and
\begin{equation}
 \hat{a}_+ f_{a,m} = \left\{\begin{matrix}
                                  {\rm i} f_{a-1,m+1}  & \textrm{if } m<0 \\
				  {\rm i} f_{a,m+1}  & \textrm{if } m\geq0
                                 \end{matrix}  \right..
\end{equation}
\end{subequations}
Using the wave function ansatz given by Eq.~(\ref{eq:wavefunc}) for the two block diagonal parts of the Hamiltonian $H_K$ (with $\lambda_R=0$) the Schr\"odinger equation reduces to two simple matrix eigenproblems. 
The solutions of this eigenproblem are the energy eigenvalues given by Eqs.~(\ref{eq:E_B}) and (\ref{eq:E_A}), as well as the $\Lambda_{a,m,s}^{\pm} = {\rm i}\hbar\omega_c\left(E^{\pm}_{a,m,s}+s\Delta_{SO}\right)^{-1}$ amplitudes in the Eq. (\ref{eq:wavefunc}) wave function ansatz.

\emph{Quantum dot:}
For the QD system we follow the same procedure as described above.
The effect of the operators $\hat{a}_{\pm}$ on the basis functions (\ref{eq:basis_antidot}) reads
\begin{subequations}
\begin{equation}
 \hat{a}_- f_{a,m} = \left\{\begin{matrix}
                                  \frac{-{\rm i}a}{|m|+1} f_{a+1,m-1}  & \textrm{if } m\leq0 \\
				  -{\rm i}m f_{a,m-1}  & \textrm{if } m>0
                                 \end{matrix}  \right.,
\end{equation}
and
\begin{equation}
 \hat{a}_+ f_{a,m} = \left\{\begin{matrix}
                                  {\rm i}m f_{a-1,m+1} & \textrm{if } m<0 \\
				  \frac{a-|m|-1}{{\rm i}(|m|+1)} f_{a,m+1}  & \textrm{if } m\geq0
                                 \end{matrix}  \right..
\end{equation}
\end{subequations}
The solutions of the eigenproblem are given by the energy eigenvalues of Eqs.~(\ref{eq:E_B}) and (\ref{eq:E_A}) and by the $\Lambda_{a,m,s}^{\pm}$ amplitude that is
\begin{equation}
 \Lambda_{a,m,s}^{\pm} = \left\{\begin{matrix}
			  {\rm i}m\hbar\omega_c\;(E^{\pm}_{a,m,s}+s\Delta_{SO})^{-1} & \textrm{if } m<0, \\
                          {\rm i}\frac{E^{\pm}_{a,m,s}-s\Delta_{SO}}{(m+1)\hbar\omega_c} & \textrm{if } m\geq0.
                         \end{matrix} \right.
\end{equation}

\end{document}